# Micromechanical bolometers for sub-Terahertz detection at room temperature.

Leonardo Vicarelli[1], Alessandro Tredicucci[1,2], Alessandro Pitanti[1]

[1] Laboratorio NEST – Scuola Normale Superiore, and Istituto Nanoscienze – CNR, Piazza San Silvestro 12, 56127 Pisa, Italy
[2] Dipartimento di Fisica, Università di Pisa, Largo B. Pontecorvo 3, 56127 Pisa, Italy

**ABSTRACT:** Fast, room temperature imaging at THz and sub-THz frequencies is an interesting feature which could unleash the full potential of plenty applications in security, healthcare and industrial production. In this Letter we introduce micromechanical bolometers based on silicon nitride trampoline membranes as broad-range detectors, down to the sub-THz frequencies. They show, at the largest wavelengths, room-temperature noise-equivalent-powers comparable to state-of-the-art commercial devices (~100 pW Hz$^{-1/2}$); adding the good operation speed and the easy, large-scale fabrication process, the trampoline membrane could be the next candidate for cheap, room temperature THz imaging and related applications.

## INTRODUCTION

Terahertz (THz) radiation detectors have been developed for over a century[1], resulting in a large variety of solutions[2–4] devised to suit an even larger range of applications, including astronomy[5], medical diagnostics[6,7], communications[8], industrial inspection[9–11], security scanning of people[12] and packages[13]. However, a broad exploitation of THz technology is still far from being achieved, as most of the proposed devices remain at a fundamental research level, unable to breach the commercialization barrier[11,14]. Nevertheless, several user-friendly cameras[14,15] for incoherent detection of THz (0.3-10 THz) and sub-THz (0.1-0.3 THz) radiation were recently introduced in the market with a relatively affordable price (few thousand $). While R&D companies currently represent the majority of customers, a few of them were actually deployed in airports as THz body scanners[16,17], where non-contact security screening of passengers has become increasingly important. Despite the absence of spectroscopic information, an intensity-based THz image is, in fact, sufficient to identify dangerous items hidden under clothes, paper or plastic[18], without being harmful to human health[19]. It seems therefore that a small market has formed, requiring inexpensive THz cameras with low-medium resolution (320x240 pixels), operating at room temperature with video acquisition rate (30-60 Hz) and, possibly, with low noise (NEP<100 pW Hz$^{-1/2}$) in order to also passively detect human blackbody radiation[20].

One of the most frequently adopted solutions for THz detection in these commercial cameras are microbolometer-based focal plane arrays (FPA)[21–27]. Originally designed for the infrared range, they were adapted to detect THz radiation tuning the materials and structure of the absorption layer, adding an antenna and adjusting the size of their resonant cavity. Such modifications led to impressive improvements of their detection capabilities, resulting in state-of-the-art low NEP (<10 pW Hz$^{-1/2}$ and <500 pW Hz$^{-1/2}$ above and below 1 THz, respectively[15]) and demonstrating that uncooled bolometers can be very competitive in this range. From this perspective, other bolometric approaches that were successful in the infrared range could, in principle, be modified to operate at lower frequencies.

Therefore, we have focused our attention on an emerging type of bolometric detectors based on microelectromechanical (MEMS) resonators. In this kind of device, a small suspended structure (typical size 10-1000 μm) can freely vibrate, in air or vacuum, with its own characteristic resonance frequency, which can be measured by electrical or optical means. The heat generated by the absorbed radiation induces a thermal expansion of the vibrating structure, shifting its resonance frequency. Considering exclusively the infrared spectral range, several MEMS resonator bolometers have been proposed in recent years, resulting in wide assortment of geometries and materials. These include GaN plates suspended by two tethers[28], Si$_3$N$_4$ drums with thin film absorber[29], Si torsional resonator with TiN absorber[30,31], Graphene-AlN-Pt nanoplate[32], SiN phononic crystal membranes[33] and Graphene drum/trampolines[34]. Up to this date, and limited to our knowledge, only one work has demonstrated a MEMS resonator capable of room-temperature bolometric detection in the THz range[35,36]. Their device, based on doubly clamped GaAs beams coated with a 15-nm-thick NiCr THz absorbing film, reached good performance in terms of noise (optical NEP~500 pW Hz$^{-1/2}$) and speed (1 kHz bandwidth) at room temperature, with the aid of a Si hemispherical lens. Another MEMS approach based on GaAs-Au meta-atom THz resonators[37,38] exploits combined coulomb/photothermal detection; the fast device operation achieved (MHz) relies on radiation source modulation at the mechanical mode frequency.

In this work, we show the realization of an ultra-sensitive micromechanical resonator bolometer using a Si$_3$N$_4$ trampoline with a 35-nm-thick Cr-Au coating, challenging the state-of-the-art for room-temperature bolometric detectors in the sub-THz range (140 GHz), with a minimum NEP of ~100 $pW Hz^{-1/2}$ and a detection speed of 40 Hz. Additionally, we demonstrate broadband bolometric detection, covering both infrared and visible light.

The non-trivial choice of the Si$_3$N$_4$ trampoline as the most suitable platform for our bolometric detector is the result of an extensive analysis. Thanks its high Q factors (Q>$10^6$) and small mass, the Si$_3$N$_4$ trampoline has excelled in many ultrasensitive applications, such as electron-spin detection[39], position sensors[40] and displacement detectors[41] with pm resolution. Compared to micro strings/plates/beams MEMS reso-nators, the size of the large central plate can be set to match the diffraction limited area at THz frequencies, thus maximizing radiation absorption and removing the need of an additional focusing Si lens. At the same time, the small tethers' width minimizes heat losses to the surrounding Si frame, ensuring the formation of a steep temperature gradient. If confronted with the conceptually similar Graphene trampoline[34], our device performs worse in the infrared range, but offers a much simpler, economical and well-established fabrication procedure, which can be easily extended to large scale for the realization of arrays. and, possibly inexpensive, multi-pixel THz cameras.

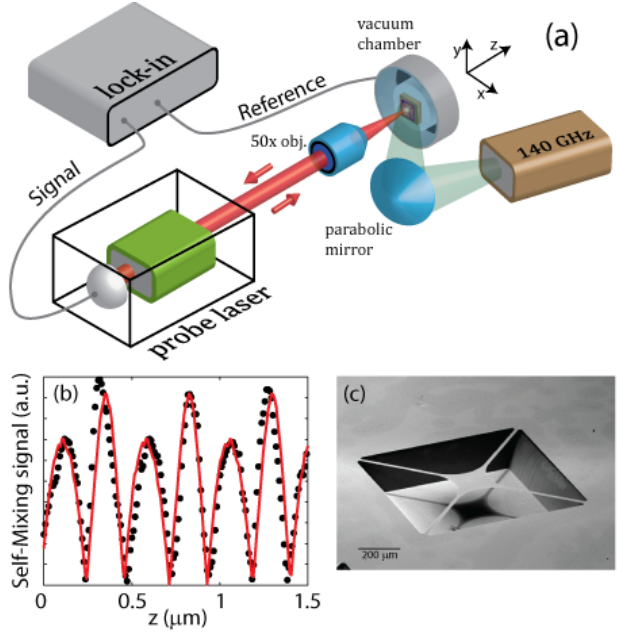

Figure 1: (a) sketch of the experimental setup. (b) SEM micrograph of a trampoline membrane (device M2). (c) Experimental (black dots) and numerical (red line) static Self-Mixing signal.

## EXPERIMENTAL SETUP

The setup we used to characterize our device is sketched in Fig. 1 (a). Similarly to other MEMS based bolometric approaches[31,33,37] and to Golay cells[42], we performed the readout of the trampoline vibration frequency by optical means. We used a self-mixing (SM) interferometric technique, with a near-infrared laser (945 nm) working as a probe; after emission and focusing on the trampoline's surface, the laser is reflected back into the cavity itself. Fluctuations of the intracavity field amplitude carry the memory of light interacting with the environment and therefore can be used to probe tiny movement of the trampoline. In our setup, this can be done by using a photodiode integrated within the laser cavity and used to probe the laser power. The dynamics of SM is well described through the Lang-Kobayashi equations[43], which, in steady state conditions, can be reduced to a compact expression for the laser emission frequency, $\omega_{SM}$, the cavity carrier density $n_s$ and the photon density $P_s$ [44]:

$$\omega_0 - \omega_{SM} = \frac{k}{\tau_c}\sqrt{1+\alpha^2}\sin(\omega_{SM}\tau_{ext} + \arctan\alpha) \quad (1)$$

$$n_s = n_{th} - \frac{2k}{G_n\tau_c}\cos(\omega_{SM}\tau_{ext}) \quad (2)$$

$$P_s = |E_s|^2 = \frac{1}{G_n(n_s-n_0)}\left(R - \frac{n_s}{\tau_s}\right) \quad (3)$$

where $\omega_0$ is the laser frequency without feedback, $\alpha$ is the linewidth enhancement factor, $k$ a matching constant, $G_n$ the modal gain factor and $R$ the carrier injection rate. $n_{th}$ and $n_0$ are the carrier densities at threshold and transparency, respectively. $\tau_c$, $\tau_s$ and $\tau_{ext}$ are the cavity round-trip time, the carrier lifetime and the feedback time, respectively, the latter representing the time necessary for the photons to return into the cavity after emission. A small vibration of the reflecting target translates in a change in $\tau_{ext}$, changing the solution of eq. (1), (2) and (3). As a preliminary characterization experiment, we employed a fixed mirror as a target, mounted on a piezoelectric actuator and on a motorized stage. Using a lock-in amplifier, we applied a reference sinusoidal voltage to the piezo to induce small oscillations (few tens of nm) of the mirror along the z-axis, then acquired the demodulated SM signal at different distances (z-axis) from the laser itself, moving the motorized stage. The resulting signal amplitude is shown as black dots in Fig. 1 (b). The experimental data are in good agreement with numerical analysis based on the aforementioned equations, where the parameters used are compatible with values reported in the literature[41]. After calibration, we switched the fixed mirror with the $Si_3N_4$ trampoline microresonators.

We fabricated and characterized two different membranes starting from a 300-nm-thick $Si_3N_4$ film: the first one (M1) has a 300x300 μm central plate and 490x20 μm tethers, while the second (M2) has a smaller 200x200 μm plate and longer 560x20 μm tethers. Both membranes are clamped to a square Si supporting frame, 1x1 mm in size. A SEM micrograph of device M2 is shown in Fig. 1 (c). We metalized both resonators with 5/30 nm Cr/Au in order to increase the material absorption and the overall thermal conductivity. More details on the device geometry and fabrication are reported in the Methods section.

## INFRARED BOLOMETRY

Starting from device M1, we placed the microresonator on a piezoelectric actuator inside a vacuum chamber with a transparent window in order to grant light access (both visible, infrared and THz). We shined the laser light at the trampoline center and used a sinusoidal voltage as bias for the piezo actuator as well as reference for the lock-in demodulation, similarly to the fixed mirror configuration. When the voltage frequency was resonant with one of the membrane modes, a clear signal appeared in the lock-in amplifier, as can be seen in Fig. 2 (a). The demodulated amplitude has a Lorentzian shape, with a corresponding phase slip of $\pi$ when crossing the resonance frequency, located approximately at 80 kHz for this particular membrane. We identified this mode as the fundamental drum mode of the mechanical resonator, which is characterized by a rigid shift of the central plate perpendicularly to the membrane plane. The map shown in Fig. 2 (d) agrees with our observation, showing a constant demodulated signal across the membrane center, with a decreasing signal running around the tethers. We estimated the quality factor of such mode from best-fit analysis of the Lorentzian curve, resulting in Q ~900.

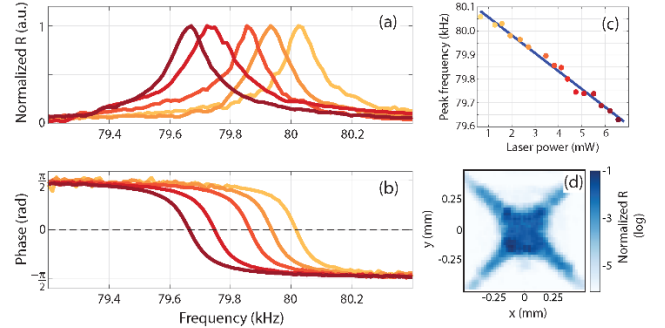

Figure 2: Device M1. Demodulated amplitude (a) and phase (b) when probing the trampoline resonator at different laser powers. (c) shift of the resonant frequency as a function of laser power. (d) Map of the demodulated amplitude around the membrane center.

Interestingly, the resonance peak position depends linearly on the impinging power of the SM infrared laser. As can be seen in the Fig. 2 (c), as well as in the set of curves in panels (a) e (b), increasing laser power red-shifts the resonance with a slope of 75 kHz/W, corresponding to a normalized responsivity of 0.94 W$^{-1}$ (calculated using the resonance frequency at zero incident power $f_0$=80.1 kHz). This is due to an increase of the device temperature induced by the laser heating.

In order to explore the dependency of responsivity on the trampoline geometry, we repeated the characterization on device M2. As the beam spot size of the infrared laser is considerably smaller (36 μm) than the central plate size (200-300 μm), the amount of absorbed radiation is the same for both M1 and M2, but the latter has longer tethers, resulting in a lower thermal conductance and therefore, higher responsivity. This observation was indeed confirmed by a measured responsivity of 187 kHz/W for device M2, resulting in a 2.05 W$^{-1}$ normalized responsivity (Figure S1 in the Supp. Info). To better understand the observed values, we performed Finite-Element-Method (FEM) simulations of the two devices and compared the results (see Methods for details about the simulation). The simulated infrared responsivities were 560 kHz/W and 620 kHz/W for device M1 and M2, respectively, giving a qualitative agreement with the measured values. The fact that the simulated values are 3 to 7 times higher than the measured ones, could be attributed to small defects in actual membrane geometry and uncertainties on the thin $Si_3N_4$ and Cr/Au film empirical parameters, such as thermal conductivity, intrinsic stress and thermal expansion coefficient, which are strongly dependent on thickness (or the amount Si content for the $Si_3N_4$) and can be very different from bulk values[45–48].

## BROADBAND CHARACTERIZATION

After this initial characterization in the infrared range, we continued exploring the bolometric effect of our thermomechanical device using an external source illumination, while keeping the probe laser as weak as possible (1.5 mW, slightly above the lasing threshold). To show the broadband operation of our device, we used a green laser diode (563 THz, or 532 nm) and a sub-THz microwave source (0.14 THz). Both sources have been mounted following the scheme of Fig. 1 (a), employing a parabolic mirror to focus the emission on the trampoline membrane. Despite the focusing, the spot size of the sub-THz beam in the detector plane was approximately 7.1x4.3 mm, which is significantly larger than the device. For this reason, the sub-THz responsivities have been calculated using the incident power inside the diffraction limited area $\lambda^2/4$~1.15 mm$^2$, which is equal to 0.32 mW. The green laser, instead, was entirely focused inside the trampoline's central plate (spot size 108 μm), but it's intensity was reduced with a 2.1 neutral density (ND) filter, resulting in a total 0.016 mW incident power.

In order to measure the responsivities with such low incident powers, we modified the detection experiment to operate in an open-loop configuration, fixing the excitation frequency of the piezo actuator and monitoring the temperature induced phase shift of the demodulated signal. As visual guidance for the reader, this corresponds to the point of maximum slope in Fig. 2(b). The phase responsivities measured on the two membranes, employing both sources, are reported in Fig. 3. The measurements have been performed with modulated sources in order to assess the operational speed of the thermomechanical bolometers. The maximum values, reached at low modulation frequencies, are 350 kHz/W (norm. 4.4 W$^{-1}$) and 140 kHz/W (norm. 1.6 W$^{-1}$) with the sub-THz source for membranes M1 and M2, respectively, while the responsivity for the green laser tops at 560 kHz/W (norm. 6.2 W$^{-1}$) on device M2. The green laser responsivities of both devices were also measured at zero modulation frequency, in a similar configuration to the infrared case, tuning the green laser power with ND filters of increasing opacity. The measured responsivities are 520 kHz/W and 600 kHz/W for device M1 and M2, respectively, in agreement with the values obtained looking at the phase variation (measurements are reported in the Supp. Info, Figure S2 and S3).

Looking at the modulation frequency dependence, both devices show qualitatively similar trends, with a 3 dB cut-off frequency of about 40 Hz and 20 Hz, for the sub-THz and visible source, respectively. The bottleneck here is represented by the thermal dynamics; the main source of dissipation for the membrane is through the tethers. By changing the tethers geometry one can expect to be able to increase the dissipation and device operational speed, at the expense of a possible reduction in the responsivity. The slight difference in the results from the two sources illuminating device M2 can be ascribed to the very different spot sizes, as described earlier in this text. In fact, the broad illumination from the sub-THz source implies that the incident power is dissipated faster, having a shorter average path to the thermal heat sink.

Another consequence of the larger illumination area is that the sub-THz responsivity of device M1 is always higher than device M2, opposite to what we observed with the infrared laser. Here the straightforward explanation comes from the larger surface of device M1 respect to M2, implying a higher total absorbed radiation. To be more specific, the approximate 2.2 ratio between the areas of the central plates is compatible with the 2.5 ratio between the measured responsivities. Examples of detected signals in time domain are reported in Fig. 3 (b) and (c) for device M1 illuminating at 0.14 THz. As can be seen, the detected signal (blue traces) perfectly reproduces the modulated source input (orange traces) for 1 Hz modulation; despite the presence of some deformations at 210 Hz modulation, the detected signal clearly follows the source dynamics. We repeated the FEM simulations with the sub-THz source and obtained 200 kHz/W (M1) and 130 kHz/W (M2), close to the measured absolute values but with a 1.5 ratio which is smaller than the experimental one. Lastly, the simulated green laser responsivity resulted in a much higher value of 2700 kHz/W (M2), due to the combination of higher absorption (17%, see Methods) and focused beam.

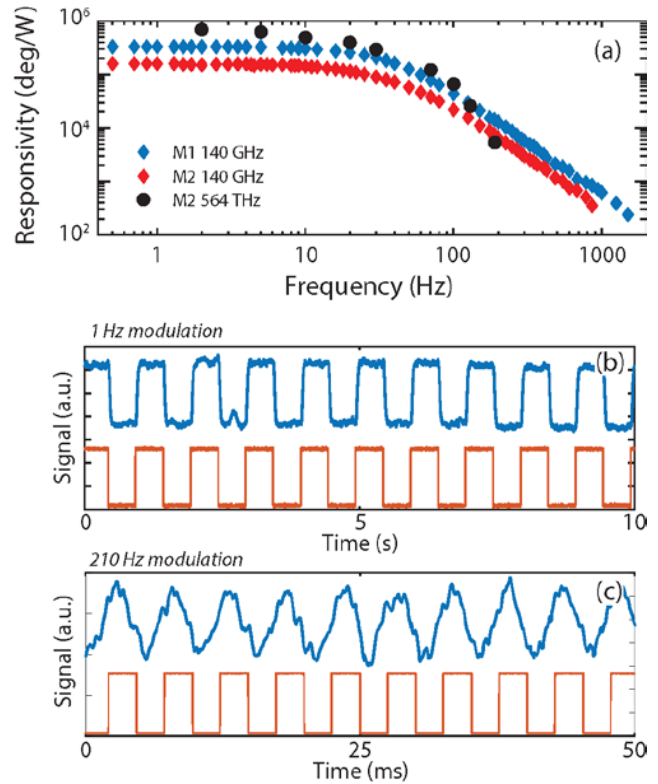

Figure 3: (a) Responsivity for the two different membranes, as a function of modulation frequency for visible (564 THz) and sub-THz (0.14 THz) ra-

diation. (b) and (c) Time plot of the phase signal for membrane M1, illuminated by the 0.14 THz source, for 1 Hz and 210 Hz modulation frequencies, respectively (blue=signal, orange=TTL modulation)

## NOISE MEASUREMENTS

To better estimate the thermomechanical bolometer performances, we considered the noise equivalent power (NEP), which represents the incident power producing a unitary detected signal-to-noise ratio. This has been obtained by taking the ratio of the measured spectral noise density and the responsivity reported in Fig. 3 (a). Figure 4 shows the NEP of devices M1 and M2 as a function of the modulation frequency, using the 0.14 THz source. Both devices share a similar trend, but device M1 consistently shows a lower NEP compared to device M2, thanks to its higher responsivity.

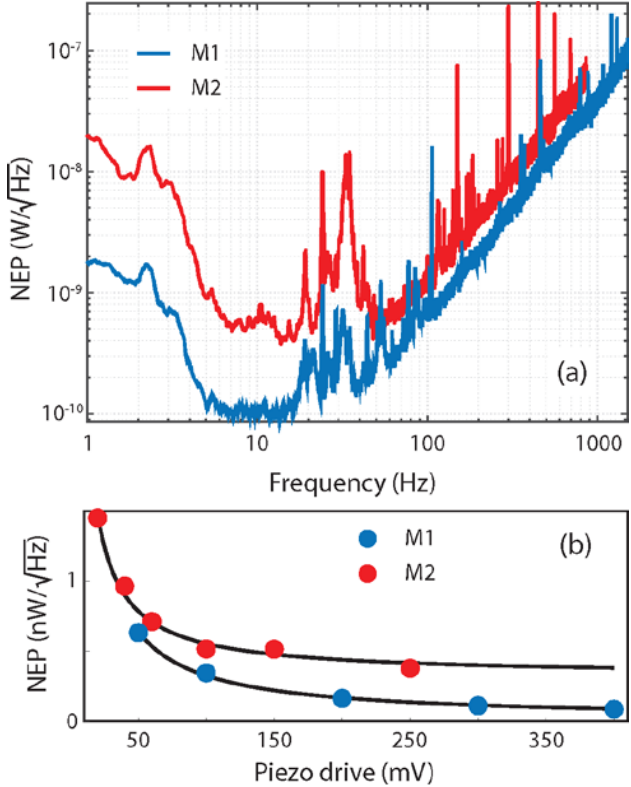

Figure 4: (a) NEP of the two membranes illuminated by the 0.14 THz source, as function of the modulation frequency. (b) Minimum NEP as function of the amplitude of the sinusoidal driving voltage applied to the piezoelectric actuator.

In the 7-14 Hz modulation interval, where the curve is mostly flat, our best device reaches a minimum NEP of $\sim 100\ pWHz^{-1/2}$. Considering the detectors operating at room temperature in the THz and sub-THz range, this NEP value is competing with the state-of-the-art technology, represented by Vanadium oxide microbolometers [21-25], Schottky diodes [49], NMOS detectors [50,51], Dyakonov-Shur plasma wave detectors [52-55] and many other photodetectors based on 2D crystals [56-60].
Below 7 Hz modulation, the NEP rapidly increases due to a combination of $1/f$ noise, thermal drift [61-63] and adsorption-desorption noise [63]. Instabilities of the SM optical read-out also contribute to the low frequency noise. In fact, air flow and small drifts of the various optical elements will change the optical path length, inducing seldom mode-hopping in the LD [64]. Going to higher frequencies, the NEP increases above 14 Hz due to the declining responsivity, while the noise remains white. On top of this general trend, several resonance peaks are visible across the whole spectrum, which are related to environmental vibrations.

NEP was also found to be strongly dependent on the amplitude of the membrane's vibration, which can be adjusted via the sinusoidal driving voltage sent to the piezoelectric actuator. In particular, the minimum NEP (in the 7-14 Hz region) is inversely proportional to the driving voltage, as shown in Fig. 4 (b), and tends to decrease at higher voltages. This behavior is related to the fact that the SM signal, and therefore the vibration amplitude, both increase linearly with the driving voltage (see Supp. Info, Figure S6), while noise remains constant. This trend remains valid until the vibration amplitude saturates, and nonlinear effects come into play, which in our case occurs above 450 mV piezo driving voltage. Interestingly, the piezo drive represents the main source of power needed to operate the device. Considering a 150 mV drive, which gives a $\sim 200\ pWHz^{-1/2}$ NEP, a single thermomechanical bolometer dissipates a power of 225 nW. This relates well to the possibilities of creating a grid of resonators, which can be used for sub-THz imaging and similar applications.

## CONCLUSION AND OUTLOOK

In conclusion, we have shown how $Si_3N_4$ trampoline membrane resonators can be an optimal platform for thermomechanical bolometric detection at room temperature in a broad range, down to the sub-THz frequencies. Direct detection experiments have shown NEP comparable with state-of-the-art, commercial devices. To increase the device responsivity in the THz range, absorption could be enhanced by tailoring the surface of the trampoline resonator with different materials and geometrical patterns. For example vanadium oxide metasurfaces[65] and metal-dielectric-metal structures[66] have been reported to absorb more than 90% radiation around 1 THz. In order to make the detector more suitable for array implementation, the self-mixing optical readout could be replaced with a magnetomotive force[40] actuator/sensor. These improvements, together with the simple and well-established fabrication procedures, would pave the way for large-scale fabrication of arrays of detectors, making the investigated platform extremely appealing for THz and sub-THz spatially resolved detection and cameras.

## METHODS

**Sample Fabrication.** The trampolines were fabricated starting from a 300 nm thick, high stress (~900 MPa), stoichiometric LPCVD $Si_3N_4$ film, grown on top of a 250 µm thick Si wafer. The trampoline shape was patterned with optical lithography (DMO MicroWriter ML3), carefully smoothing all sharp corners to reduce stress in critical points, such as the tethers' clamping to the Si frame. The unmasked $Si_3N_4$ was removed by a Reactive Ion Etching (RIE) step with CF4 and H2 gas mixture.
Then the trampolines were released with hot KOH etching solution (30% concentration). Finally, a 5/30 nm Cr/Au metallic layer was thermally evaporated on top of the whole device surface.

**Experimental Setup.**
The position of the various elements in the setup is shown in Figure 1. The Si chip containing the $Si_3N_4$ membranes was mounted inside a small vacuum chamber ($2 \times 10^{-3}$ mbar), sealed with a cyclic-olefin copolymer (COC) window, transparent to both visible and THz radiation (87.7% transmission in the infrared and visible range, 80% in the sub-THz range). The whole chamber could be moved with step-motors in the x-y plane, orthogonally to the optical beam axis. Inside the chamber, the Si chip was glued on top of a piezo actuator, used to excite the membrane vibrations, which was then attached with double-sided tape to the chamber itself. A Littrow External Cavity Diode Laser (ECDL), model DL100-L from Toptica, 945 nm wavelength, was used to sense the membrane displacement via self-mixing interferometric technique. The ECDL beam was focused on the membrane with an objective lens (Mi-tutoyo M Plan Apo 10x/0.28, focal length 200 mm), resulting in an approximate beam radius of ~36 µm in the focal plane (measured by the knife-edge method). The output of the photodiode mounted on the backside facet of the ECDL, measuring the fluctuations of the laser intensity due to self-mixing, was sent to a lock-in amplifier (Zurich Instruments MFLI). The Lorentzian curves shown in Fig 2(a) were generated with sequential frequency sweeps of the sinusoidal voltage sent to piezo actuator, using only one channel of the lock-in amplifier.

The radiation generated by the green laser and sub-THz source was focused on the membrane using an off-axis parabolic mirror (gold-coated, 2" focal length). The 0.14 THz beam was generated by a commercially available source (TeraSense Group) based on an Impact ionization Avalanche Transit Time (IMPATT) diode, equipped with a conical horn antenna. The 564 THz ($\lambda$=532 nm) source was a commercial green laser pointer, attenuated with an additional 2.1 ND filter in order to keep a lower incident power on the trampoline. With a calibrated pyrometer we measured the output power of both sources after the reflection in the parabolic mirror and the COC window, resulting in 10.6 mW for the sub-THz beam and 2.0 mW for the green laser, the latter reduced to 0.016 mW after the ND filter. In order to measure the responsivity of the membrane-detector as function of the source modulation frequency, a mechanical chopper (maximum 200 Hz rotation) and a TTL signal were used for the green laser and the sub-THz source, respectively. The responsivity, expressed in $deg/W$, was measured using two separate lock-in amplifiers in cascade (Zurich Instruments MFLI and Stanford Research SR830). The membrane was excited at a fixed frequency, set at the middle point between the resonance measured with and without radiation, and the phase output of the first lock-in was sent to a second lock-in amplifier, which used the chopper (or TTL) as reference. This configuration was chosen in order to maximize the phase variation respect to the incident power. However, with this configuration the linearity of the response is guaranteed only for small incident powers, where the slope of the phase plot is steepest (see Figure 2b).

**FEM simulations.**
We performed the Finite Element simulations with COMSOL Multiphysics v 5.5. The 2D trampoline shape was meshed with a free triangular mesh, then swept vertically to mesh the 3D structure. The Cr/Au layer was simulated as a single 35-nm-thick Au element. The mechanics and heating were simulated simultaneously, using the "thermal expansion" multiphysics coupling. The heating due to the lasers and THz source was simulated with the "Deposited Beam Power" option, using a Gaussian beam profile. A "time-dependent" study simulated the initial temperature spatial profile, followed by a "pre-stressed eigenfrequency" study to extract the frequency of the fundamental drum mode. The material parameters used for $Si_3N_4$ and Au are listed in Table S1 in the Supporting Information. The absorption of Cr/Au metallic layer was estimated from the real and imaginary parts of the refractive indices, and calculated to be 14% for the 0.14 THz radiation[67], 3.3% for the 945 nm infrared laser[68] and 17% for the 532 nm green laser[68]. $Si_3N_4$ was assumed to be completely transparent for all frequencies under consideration[69].

## ACKNOWLEDGMENT
This project has received funding from the EU ATTRACT project, Grant Agreement 777222.

## REFERENCES

(1) Sizov, F. F. Brief History of THz and IR Technologies. *Semicond. Physics, Quantum Electron. Optoelectron.* **2019**, *22* (1), 67–79.

(2) Lewis, R. A. A Review of Terahertz Detectors. *J. Phys. D. Appl. Phys.* **2019**, *52* (43), 433001.

(3) Sizov, F. Terahertz Radiation Detectors: The State-of-the-Art. *Semicond. Sci. Technol.* **2018**, *33* (12), 123001.

(4) Dhillon, S. S.; Vitiello, M. S.; Linfield, E. H.; Davies, A. G.; Hoffmann, M. C.; Booske, J.; Paoloni, C.; Gensch, M.; Weightman, P.; Williams, G. P.; Castro-Camus, E.; Cumming, D. R. S.; Simoens, F.; Escorcia-Carranza, I.; Grant, J.; Lucyszyn, S.; Kuwata-Gonokami, M.; Konishi, K.; Koch, M.; Schmuttenmaer, C. A.; Cocker, T. L.; Huber, R.; Markelz, A. G.; Taylor, Z. D.; Wallace, V. P.; Axel Zeitler, J.; Sibik, J.; Korter, T. M.; Ellison, B.; Rea, S.; Goldsmith, P.; Cooper, K. B.; Appleby, R.; Pardo, D.; Huggard, P. G.; Krozer, V.; Shams, H.; Fice, M.; Renaud, C.; Seeds, A.; Stöhr, A.; Naftaly, M.; Ridler, N.; Clarke, R.; Cunningham, J. E.; Johnston, M. B. The 2017 Terahertz Science and Technology Roadmap. *J. Phys. D. Appl. Phys.* **2017**, *50* (4), 043001.

(5) Walker, C. K. *Terahertz Astronomy*; CRC Press, 2016.

(6) Yu, L.; Hao, L.; Meiqiong, T.; Jiaoqi, H.; Wei, L.; Jinying, D.; Xueping, C.; Weiling, F.; Yang, Z. The Medical Application of Terahertz Technology in Non-Invasive Detection of Cells and Tissues: Opportunities and Challenges. *RSC Adv.* **2019**, *9* (17), 9354–9363.

(7) Zaytsev, K. I.; Dolganova, I. N.; Chernomyrdin, N. V; Katyba, G. M.; Gavdush, A. A.; Cherkasova, O. P.; Komandin, G. A.; Shchedrina, M. A.; Khodan, A. N.; Ponomarev, D. S.; Reshetov, I. V; Karasik, V. E.; Skorobogatiy, M.; Kurlov, V. N.; Tuchin, V. V. The Progress and Perspectives of Terahertz Technology for Diagnosis of Neoplasms: A Review. *J. Opt.* **2020**, *22* (1), 013001.

(8) Koenig, S.; Lopez-Diaz, D.; Antes, J.; Boes, F.; Henneberger, R.; Leuther, A.; Tessmann, A.; Schmogrow, R.; Hillerkuss, D.; Palmer, R.; Zwick, T.; Koos, C.; Freude, W.; Ambacher, O.; Leuthold, J.; Kallfass, I. Wireless Sub-THz Communication System with High Data Rate. *Nat. Photonics* **2013**, *7* (12), 977–981.

(9) Stecher, M.; Jördens, C.; Krumbholz, N.; Jansen, C.; Scheller, M.; Wilk, R.; Peters, O.; Scherger, B.; Ewers, B.; Koch, M. Towards Industrial Inspection with THz Systems. In *Springer Series in Optical Sciences*; 2016; Vol. 195, pp 311–335.

(10) Tao, Y. H.; Fitzgerald, A. J.; Wallace, V. P. Non-Contact, Non-Destructive Testing in Various Industrial Sectors with Terahertz Technology. *Sensors* **2020**, *20* (3), 712.

(11) Naftaly; Vieweg; Deninger. Industrial Applications of Terahertz Sensing: State of Play. *Sensors* **2019**, *19* (19), 4203.

(12) Tzydynzhapov, G.; Gusikhin, P.; Muravev, V.; Dremin, A.; Nefyodov, Y.; Kukushkin, I. New Real-Time Sub-Terahertz Security Body Scanner. *J. Infrared, Millimeter, Terahertz Waves* **2020**, *41* (6), 632–641.

(13) Shchepetilnikov, A. V.; Gusikhin, P. A.; Muravev, V. M.; Tsydynzhapov, G. E.; Nefyodov, Y. A.; Dremin, A. A.; Kukushkin, I. V. New Ultra-Fast Sub-Terahertz Linear Scanner for Postal Security Screening. *J. Infrared, Millimeter, Terahertz Waves* **2020**, *41* (6), 655–664.

(14) Simoens, F. Buyer's Guide for a Terahertz (THz) Camera. *Photoniques*. 2018, pp 58–62.

(15) Oda, N. Technology Trend in Real-Time, Uncooled Image Sensors for Sub-THz and THz Wave Detection. In *Micro- and Nanotechnology Sensors, Systems, and Applications VIII*; George, T., Dutta, A. K., Islam, M. S., Eds.; 2016; Vol. 9836, p 98362P.

(16) MM-imager - MC2 Technologies https://www.mc2-

(17) TS4-SC - Thruvision Ltd. https://thruvision.com/products/customs-and-border-security-cameras/.

(18) Federici, J. F.; Schulkin, B.; Huang, F.; Gary, D.; Barat, R.; Oliveira, F.; Zimdars, D. THz Imaging and Sensing for Security Applications—Explosives, Weapons and Drugs. *Semicond. Sci. Technol.* **2005**, *20* (7), S266–S280.

(19) Kleine-Ostmann, T.; Jastrow, C.; Baaske, K.; Heinen, B.; Schwerdtfeger, M.; Karst, U.; Hintzsche, H.; Stopper, H.; Koch, M.; Schrader, T. Field Exposure and Dosimetry in the THz Frequency Range. *IEEE Trans. Terahertz Sci. Technol.* **2014**, *4* (1), 12–25.

(20) Čibiraitė-Lukenskienė, D.; Ikamas, K.; Lisauskas, T.; Krozer, V.; Roskos, H. G.; Lisauskas, A. Passive Detection and Imaging of Human Body Radiation Using an Uncooled Field-Effect Transistor-Based THz Detector. *Sensors* **2020**, *20* (15), 4087.

(21) Oda, N.; Kurashina, S.; Miyoshi, M.; Doi, K.; Ishi, T.; Sudou, T.; Morimoto, T.; Goto, H.; Sasaki, T. Microbolometer Terahertz Focal Plane Array and Camera with Improved Sensitivity in the Sub-Terahertz Region. *J. Infrared, Millimeter, Terahertz Waves* **2015**, *36* (10), 947–960.

(22) Dufour, D.; Marchese, L.; Terroux, M.; Oulachgar, H.; Généreux, F.; Doucet, M.; Mercier, L.; Tremblay, B.; Alain, C.; Beaupré, P.; Blanchard, N.; Bolduc, M.; Chevalier, C.; D'Amato, D.; Desroches, Y.; Duchesne, F.; Gagnon, L.; Ilias, S.; Jerominek, H.; Lagacé, F.; Lambert, J.; Lamontagne, F.; Le Noc, L.; Martel, A.; Pancrati, O.; Paultre, J.-E.; Pope, T.; Provençal, F.; Topart, P.; Vachon, C.; Verreault, S.; Bergeron, A. Review of Terahertz Technology Development at INO. *J. Infrared, Millimeter, Terahertz Waves* **2015**, *36* (10), 922–946.

(23) Simoens, F.; Meilhan, J. Terahertz Real-Time Imaging Uncooled Array Based on Antenna- and Cavity-Coupled Bolometers. *Philos. Trans. R. Soc. A Math. Phys. Eng. Sci.* **2014**, *372* (2012), 20130111.

(24) Jang, D.; Kimbrue, M.; Yoo, Y.-J.; Kim, K.-Y. Spectral Characterization of a Microbolometer Focal Plane Array at Terahertz Frequencies. *IEEE Trans. Terahertz Sci. Technol.* **2019**, *9* (2), 150–154.

(25) Romano, M.; Chulkov, A.; Sommier, A.; Balageas, D.; Vavilov, V.; Batsale, J. C.; Pradere, C. Broadband Sub-Terahertz Camera Based on Photothermal Conversion and IR Thermography. *J. Infrared, Millimeter, Terahertz Waves* **2016**, *37* (5), 448–461.

(26) Oden, J.; Meilhan, J.; Lalanne-Dera, J.; Roux, J.-F.; Garet, F.; Coutaz, J.-L.; Simoens, F. Imaging of Broadband Terahertz Beams Using an Array of Antenna-Coupled Microbolometers Operating at Room Temperature. *Opt. Express* **2013**, *21* (4), 4817.

(27) Grant, J.; Escorcia-Carranza, I.; Li, C.; McCrindle, I. J. H.; Gough, J.; Cumming, D. R. S. A Monolithic Resonant Terahertz Sensor Element Comprising a Metamaterial Absorber and Micro-Bolometer. *Laser Photon. Rev.* **2013**, *7* (6), 1043–1048.

(28) Gokhale, V. J.; Rais-Zadeh, M. Uncooled Infrared Detectors Using Gallium Nitride on Silicon Micromechanical Resonators. *J. Microelectromechanical Syst.* **2014**, *23* (4), 803–810.

(29) Piller, M.; Luhmann, N.; Chien, M.-H.; Schmid, S. Nanoelectromechanical Infrared Detector. In *Optical Sensing, Imaging, and Photon Counting: From X-Rays to THz 2019*; Mitrofanov, O., Ed.; SPIE, 2019; Vol. 1108802, p 1.

(30) Laurent, L.; Yon, J.-J.; Moulet, J.-S.; Roukes, M.; Duraffourg, L. 12-Mm-Pitch Electromechanical Resonator for Thermal Sensing. *Phys. Rev. Appl.* **2018**, *9* (2), 024016.

(31) Zhang, X. C.; Myers, E. B.; Sader, J. E.; Roukes, M. L. Nanomechanical Torsional Resonators for Frequency-Shift Infrared Thermal Sensing. *Nano Lett.* **2013**, *13* (4), 1528–1534.

(32) Qian, Z.; Hui, Y.; Liu, F.; Kang, S.; Kar, S.; Rinaldi, M. Graphene–Aluminum Nitride NEMS Resonant Infrared Detector. *Microsystems Nanoeng.* **2016**, *2* (1), 16026.

(33) Sadeghi, P.; Tanzer, M.; Luhmann, N.; Piller, M.; Chien, M.-H.; Schmid, S. Thermal Transport and Frequency Response of Localized Modes on Low-Stress Nanomechanical Silicon Nitride Drums Featuring a Phononic-Band-Gap Structure. *Phys. Rev. Appl.* **2020**, *14* (2), 024068.

(34) Blaikie, A.; Miller, D.; Alemán, B. J. A Fast and Sensitive Room-Temperature Graphene Nanomechanical Bolometer. *Nat. Commun.* **2019**, *10* (1), 4726.

(35) Zhang, Y.; Hosono, S.; Nagai, N.; Song, S. H.; Hirakawa, K. Fast and Sensitive Bolometric Terahertz Detection at Room Temperature through Thermomechanical Transduction. *J. Appl. Phys.* **2019**, *125* (15).

(36) Morohashi, I.; Zhang, Y.; Qiu, B.; Irimajiri, Y.; Sekine, N.; Hirakawa, K.; Hosako, I. Rapid Scan THz Imaging Using MEMS Bolometers. *J. Infrared, Millimeter, Terahertz Waves* **2020**, *41* (6), 675–684.

(37) Belacel, C.; Todorov, Y.; Barbieri, S.; Gacemi, D.; Favero, I.; Sirtori, C. Optomechanical Terahertz Detection with Single Meta-Atom Resonator. *Nat. Commun.* **2017**, *8* (1), 1578.

(38) Calabrese, A.; Gacemi, D.; Jeannin, M.; Suffit, S.; Vasanelli, A.; Sirtori, C.; Todorov, Y. Coulomb Forces in THz Electromechanical Meta-Atoms. *Nanophotonics* **2019**, *8* (12), 2269–2277.

(39) Fischer, R.; McNally, D. P.; Reetz, C.; Assumpção, G. G. T.; Knief, T.; Lin, Y.; Regal, C. A. Spin Detection with a Micromechanical Trampoline: Towards Magnetic Resonance Microscopy Harnessing Cavity Optomechanics. *New J. Phys.* **2019**, *21* (4), 043049.

(40) Chien, M.-H.; Steurer, J.; Sadeghi, P.; Cazier, N.; Schmid, S. Nanoelectromechanical Position-Sensitive Detector with Picometer Resolution. *ACS Photonics* **2020**, *7* (8), 2197–2203.

(41) Baldacci, L.; Pitanti, A.; Masini, L.; Arcangeli, A.; Colangelo, F.; Navarro-Urrios, D.; Tredicucci, A. Thermal Noise and Optomechanical Features in the Emission of a Membrane-Coupled Compound Cavity Laser Diode. *Sci. Rep.* **2016**, *6* (1), 31489.

(42) Zahl, H. A.; Golay, M. J. E. Pneumatic Heat Detector. *Rev. Sci. Instrum.* **1946**, *17* (11), 511–515.

(43) Lang, R.; Kobayashi, K. External Optical Feedback Effects on Semiconductor Injection Laser Properties. *IEEE J. Quantum Electron.* **1980**, *16* (3), 347–355.

(44) Spencer, P.; Rees, P.; Pierce, I. Theoretical Analysis. In *Unlocking Dynamical Diversity*; John Wiley & Sons, Ltd: Chichester, UK, 2005; pp 23–54.

(45) Ftouni, H.; Blanc, C.; Tainoff, D.; Fefferman, A. D.; Defoort, M.; Lulla, K. J.; Richard, J.; Collin, E.; Bourgeois, O. Thermal Conductivity of Silicon Nitride Membranes Is Not Sensitive to Stress. *Phys. Rev. B* **2015**, *92* (12), 125439.

(46) Mag-Isa, A. E.; Jang, B.; Kim, J. H.; Lee, H. J.; Oh, C. S. Coefficient of Thermal Expansion Measurements for Freestanding Nanocrystalline Ultra-Thin Gold Films. *Int. J. Precis. Eng. Manuf.* **2014**, *15* (1), 105–110.

(47) Lugo, J. M.; Oliva, A. I. Thermal Properties of Metallic Films at Room Conditions by the Heating Slope. *J. Thermophys. Heat Transf.* **2016**, *30* (2), 452–460.

(48) Habermehl, S. Coefficient of Thermal Expansion and Biaxial Young's Modulus in Si-Rich Silicon Nitride Thin Films. *J. Vac. Sci. Technol. A Vacuum, Surfaces, Film.* **2018**, *36* (2), 021517.

(49) Hesler, J. L.; Crowe, T. W. NEP and Responsivity of THz Zero-Bias Schottky Diode Detectors. In *2007 Joint 32nd International Conference on Infrared and Millimeter Waves and the 15th International Conference on Terahertz Electronics*; IEEE, 2007; pp 844–845.

(50) Ojefors, E.; Pfeiffer, U. R.; Lisauskas, A.; Roskos, H. G. A 0.65 THz Focal-Plane Array in a Quarter-Micron CMOS Process Technology. *IEEE J. Solid-State Circuits* **2009**, *44* (7), 1968–1976.

(51) Pleteršek, A.; Trontelj, J. A Self-Mixing NMOS Channel-Detector Optimized for Mm-Wave and THZ Signals. *J. Infrared, Millimeter, Terahertz Waves* **2012**, *33* (6), 615–626.

(52) Bianco, F.; Perenzoni, D.; Convertino, D.; De Bonis, S. L.; Spirito, D.; Perenzoni, M.; Coletti, C.; Vitiello, M. S.; Tredicucci, A. Terahertz Detection by Epitaxial-Graphene Field-Effect-Transistors on Silicon Carbide. *Appl. Phys. Lett.* **2015**, *107* (13), 131104.

(53) Tauk, R.; Teppe, F.; Boubanga, S.; Coquillat, D.; Knap, W.; Meziani, Y. M.; Gallon, C.; Boeuf, F.; Skotnicki, T.; Fenouillet-Beranger, C.; Maude, D. K.; Rumyantsev, S.; Shur, M. S. Plasma Wave Detection of Terahertz Radiation by Silicon Field Effects Transistors: Responsivity and Noise Equivalent Power. *Appl. Phys. Lett.* **2006**, *89* (25), 253511.

(54) Bauer, M.; Venckevičius, R.; Kašalynas, I.; Boppel, S.; Mundt, M.; Minkevičius, L.; Lisauskas, A.; Valušis, G.; Krozer, V.; Roskos, H. G. Antenna-Coupled Field-Effect Transistors for Multi-Spectral Terahertz Imaging up to 425 THz. *Opt. Express* **2014**, *22* (16), 19235.

(55) Viti, L.; Hu, J.; Coquillat, D.; Knap, W.; Tredicucci, A.; Politano, A.; Vitiello, M. S. Black Phosphorus Terahertz Photodetectors. *Adv. Mater.* **2015**, *27* (37), 5567–5572.

(56) Wang, X.; Cui, Y.; Li, T.; Lei, M.; Li, J.; Wei, Z. Recent Advances in the Functional 2D Photonic and Optoelectronic Devices. *Adv. Opt. Mater.* **2019**, *7* (3), 1801274.

(57) Chen, Y.; Ma, W.; Tan, C.; Luo, M.; Zhou, W.; Yao, N.; Wang, H.; Zhang, L.; Xu, T.; Tong, T.; Zhou, Y.; Xu, Y.; Yu, C.; Shan, C.; Peng, H.; Yue, F.; Wang, P.; Huang, Z.; Hu, W. Broadband $Bi_2O_2Se$ Photodetectors from Infrared to Terahertz. *Adv. Funct. Mater.* **2021**, *2009554*, 2009554.

(58) Xu, H.; Guo, C.; Zhang, J.; Guo, W.; Kuo, C.; Lue, C. S.; Hu, W.; Wang, L.; Chen, G.; Politano, A.; Chen, X.; Lu, W. $PtTe_2$-Based Type-II Dirac Semimetal and Its van Der Waals Heterostructure for Sensitive Room Temperature Terahertz Photodetection. *Small* **2019**, *15* (52), 1903362.

(59) Guo, C.; Guo, W.; Xu, H.; Zhang, L.; Chen, G.; D'Olimpio, G.; Kuo, C.-N.; Lue, C. S.; Wang, L.; Politano, A.; Chen, X.; Lu, W. Ultrasensitive Ambient-Stable $SnSe_2$-Based Broadband Photodetectors for Room-Temperature IR/THz Energy Conversion and Imaging. *2D Mater.* **2020**, *7* (3), 035026.

(60) Rogalski, A.; Kopytko, M.; Martyniuk, P. Two-Dimensional Infrared and Terahertz Detectors: Outlook and Status. *Appl. Phys. Rev.* **2019**, *6* (2), 021316.

(61) Sansa, M.; Sage, E.; Bullard, E. C.; Gély, M.; Alava, T.; Colinet, E.; Naik, A. K.; Villanueva, L. G.; Duraffourg, L.; Roukes, M. L.; Jourdan, G.; Hentz, S. Frequency Fluctuations in Silicon Nanoresonators. *Nat. Nanotechnol.* **2016**, *11* (6), 552–558.

(62) Reinhardt, C.; Müller, T.; Bourassa, A.; Sankey, J. C. Ultralow-Noise SiN Trampoline Resonators for Sensing and Optomechanics. *Phys. Rev. X* **2016**, *6* (2), 1–8.

(63) Cleland, A. N.; Roukes, M. L. Noise Processes in Nanomechanical Resonators. *J. Appl. Phys.* **2002**, *92* (5), 2758–2769.

(64) Giuliani, G.; Norgia, M.; Donati, S.; Bosch, T. Laser Diode Self-Mixing Technique for Sensing Applications. *J. Opt. A Pure Appl. Opt.* **2002**, *4* (6), S283–S294.

(65) Song, Z.; Zhang, J. Achieving Broadband Absorption and Polarization Conversion with a Vanadium Dioxide Metasurface in the Same Terahertz Frequencies. *Opt. Express* **2020**, *28* (8), 12487.

(66) Zhang, X.; Li, H.; Wei, Z.; Qi, L. Metamaterial for Polarization-Incident Angle Independent Broadband Perfect Absorption in the Terahertz Range. *Opt. Mater. Express* **2017**, *7* (9), 3294.

(67) Zhou, D.; Parrott, E. P. J.; Paul, D. J.; Zeitler, J. A. Determination of Complex Refractive Index of Thin Metal Films from Terahertz Time-Domain Spectroscopy. *J. Appl. Phys.* **2008**, *104* (5), 053110.

(68) Windt, D. L. IMD—Software for Modeling the Optical Properties of Multilayer Films. *Comput. Phys.* **1998**, *12* (4), 360.

(69) Cataldo, G.; Beall, J. A.; Cho, H.-M.; McAndrew, B.; Niemack, M. D.; Wollack, E. J. Infrared Dielectric Properties of Low-Stress Silicon Nitride. *Opt. Lett.* **2012**, *37* (20), 4200.